\documentclass[pra,aps,twocolumn,superscriptaddress,notitlepage]{revtex4-2}
\usepackage{amssymb,amsfonts,amsmath,amstext,graphicx,hyperref,lipsum,empheq}
\usepackage{float}
\usepackage[normalem]{ulem}
\hypersetup{
	colorlinks=true,
	linkcolor=blue,
	filecolor=magenta,      
	urlcolor=cyan,
}
\usepackage{amsthm}
\usepackage{outlines}
\usepackage{appendix}

\usepackage{outlines}
\usepackage{subfigure}
\usepackage{cleveref}

\usepackage[dvipsnames]{xcolor}

\newcommand{\beq}[0]{\begin{equation}}
\newcommand{\eeq}[0]{\end{equation}}
\newcommand{\non}{\nonumber}
\def\be{\begin{equation}}
\def\ee{\end{equation}}
\def\bea{\begin{eqnarray}}
\def\eea{\end{eqnarray}}
\newcommand{\ba}{\begin{eqnarray}}
\newcommand{\ea}{\end{eqnarray}}
\usepackage{hyperref}

\def\BraVert{\egroup\,\mid\,\bgroup}

\definecolor{myblue}{rgb}{.8, .8, 1}

\begin{document}

\title{Discrete time crystals in the presence of non-Markovian dynamics}
\author{Bandita Das}%
\affiliation{Department of Physical Sciences, Indian Institute of Science Education and Research Berhampur, Berhampur 760010, India}
\author{Noufal Jaseem }%
\affiliation{Department of Physical Sciences, Indian Institute of Science Education and Research Berhampur, Berhampur 760010, India}
\author{Victor Mukherjee}%
\affiliation{Department of Physical Sciences, Indian Institute of Science Education and Research Berhampur, Berhampur 760010, India}

\date{\today}

\begin{abstract}
We study discrete time crystals (DTCs) in periodically driven quantum systems, in the presence of non-Markovian dissipation. In contrast to DTCs observed in earlier works in the presence of Markovian dynamics, using the open Dicke model in presence of Jaynes-Cummings-like dissipation, we show that non-Markovian regime can be highly beneficial for stabilizing DTCs over a wide range of parameter values. This may be attributed to periodically varying dissipation rates even at long times in the case of non-Markovian dynamics. Further the Markovian and non-Markovian regimes show sharp distinctions for intermediate strengths of the dissipator coefficient, with a time-independent steady-state in the Markovian regime being replaced by varied dynamical phases, including DTC order, in the non-Markovian regime. We also verify the robustness of the DTC phase in the non-Markovian regime by introducing errors both in the Hamiltonian as well as in the dissipation.  Our study shows the possibility of using DTC as a probe for non-Markovian dynamics in periodically modulated open quantum systems, at long times.

\end{abstract}

\maketitle

\section{Introduction}
\label{secI}

Many-body quantum systems driven out of equilibrium present exciting fields of research \cite{mukherjee2024promises}, owing to the different non-trivial behaviors exhibited by such systems, such as Kibble-Zurek mechanism \cite{polkovnikov05universal, zurek05dynamics}, dynamical localization \cite{roy15fate, yousefjani23floquet}, and time-translational symmetry breaking \cite{sacha2015, zaletel23colloquium}, to name a few. Recent developments in quantum simulators have enabled researchers to experimentally study the behavior of many-body systems driven out of equilibrium as well \cite{zhang17a, king22coherent}. Consequently, theoretical and experimental research on dynamics of many-body systems have received a lot of attention in the recent years. In particular, time-translational symmetry breaking in the form of continuous \cite{iemini18boundary, montenegro23quantum,Kongkhambut2022, krishna23continoustc,liu2023photonic,Medejak2020isolated,Buca2022out} or discrete \cite{zaletel23colloquium,yao17discrete,gambetaa2019oqs,jara2024theory} time crystals have received wide interest from the community in the last decade, and also realised experimentally \cite{zhang17a, choi2017observation,  mi2022time, frey2022realization}

Discrete time crystals (DTCs) are associated with spontaneous symmetry breaking in time, and are formed in periodically driven many-body quantum systems.  In closed quantum systems DTCs have been studied widely, both theoretically \cite{zaletel23colloquium, keyserlingk16absolute, yao17discrete, norman18time, else20discrete} as well as experimentally \cite{zhang17a}, in many-body localized systems. More recently, DTCs have been shown to exist in periodically driven clean quantum systems as well \cite{huang18clean, russomanno17floquet}, and also in the presence of dissipation \cite{lazarides17fate, zhu19dicke, lazarides20time}. Such dissipative DTCs in periodically driven many-body open quantum systems can be viewed as quantum engines, wherein a part of the energy supplied through periodic drive flows to a cold bath, while the rest is obtained as output work \cite{carollo20nonequilibrium, zaletel23colloquium}. 
Analogous to different phases of matter, DTCs are robust to small perturbations. However, large perturbations, such as errors in the Hamiltonian \cite{gong18dtc}, or strong rates of dissipation \cite{zhu19dicke,cosme2023bridging}, can destroy a DTC. Consequently, finding scenarios which can result in stable DTCs is a fundamental question in the field of time-translational symmetry breaking in many-body systems. Furthermore, open questions remain regarding the existence and behavior of DTCs for different types of dissipative dynamics. Here we address the above two crucial issues by focusing on a periodically modulated many-body quantum system in the presence of non-Markovian dynamics. We show that suitably controlled non-Markovian dynamics can result in the generation of DTCs as well as other dynamical phases. 

Dissipative dynamics can be classified as Markovian or non-Markovian depending on the absence or presence of memory effects, respectively \cite{Breuer}. Markovian dynamics can be described by time-independent Lindblad superoperators, such that the system approaches a long-time steady-state monotonically. On the other hand, non-Markovian dynamics are associated with memory effects, which may result in the system  moving away from the steady state for some intervals of time \cite{kossakowski10non, rivas14quantum, breuer16colloquium,chen2023nonmarkovianity,mukherjee2015efficiency}. 

 In this work, we focus on an open Dicke model, and present results in the Markovian limit, following Ref.\cite{gong18dtc}. In case of Markovian dynamics, DTC phase obtained for weak dissipation is replaced by a time-independent steady-state (TISS) for intermediate or higher rates of dissipation \cite{zhu19dicke}. In order to study the dynamics in the non-Markovian regime, we focus on a  Jaynes-Cummings-like dissipator with tunable parameters, which allows one to traverse between the Markovian and the non-Markovian regimes \cite{Breuer}. We show that in contrast to the Markovian regime, the non-Markovian regime can be associated with a  stable  DTC for a wide range of parameter values, thereby significantly expanding the regime of parameters allowing the existence of time translational symmetry breaking. We present phase diagrams for  the Markovian and non-Markovian regimes; striking differences emerge in the response of a system in the presence of Markovian and non-Markovian dissipation, for intermediate strengths of the dissipator coefficient. This presents the intriguing possibility of using dissipative DTC as a probe for detecting  non-Markovianity  at long times. Furthermore, we show that the DTC phase persists in the non-Markovian regime even  in the presence of a random noise  in the dissipator.

We begin by discussing the model and dynamics Sec. \ref{secII}; we start by discussing the dynamics in a generic open quantum system in Sec. \ref{secIIA}, and then focus on the specific example of an open Dicke model in Sec. \ref{secIIB}. We consider Markovian dynamics in Sec. \ref{secIIM}, while we address the non-Markovian regime in Sec. \ref{secIInM}, and dissipation in the presence of a random noise in Sec. \ref{secRnM}. Finally we conclude in Sec. \ref{secIII}.

\section{Model and dynamics}
\label{secII}

\subsection{Dissipative discrete time crystals}
\label{secIIA}

 We consider a system $\mathcal{S}$ described by generic periodically modulated Hamiltonian $H(t) = H(t + T)$.  The system is coupled to a dissipative environment, such that the state $\rho$ of the system evolves following the master equation
\ba
\dot{\rho} = -\frac{i}{\hbar}[H(t), \rho(t)] + \kappa(t) \mathcal{L}[\rho(t)].
\label{eqLgen}
\ea
Here  $\mathcal{L}$ is a Lindblad superoperator, while $\kappa(t)$ determines the rate of thermalization with a bath, which we assume to be time-dependent in general. Complete positivity demands $\int^t_0 \kappa(t') dt' \geq 0$ for all $t$. Furthermore, in case of Markovian dynamics, $\kappa(t) \geq 0$ for all times, which ensures that the system approaches the long-time steady state at all times. As shown in \cite{gong18dtc} for a time-independent $\kappa(t) = \kappa_0$ $\forall~t$, a periodic modulation in $H(t)$ can result in a DTC in case of Markovian dynamics, for small values of $\kappa_0$. On the other hand, a large $\kappa_0$ is associated with rapid thermalization with the bath in a time scale $\tau_{\rm th}\sim ~\kappa_0^{-1}$, and a destruction of the DTC phase \cite{zhu19dicke}.  

In contrast to Markovian dynamics, $\kappa(t)$ can assume negative values for some time intervals in the case of non-Markovian dynamics \cite{kossakowski10non}. This results in the non-Markovian regime being associated with information back-flow, such that a negative $\kappa(t)$ may drive $\mathcal{S}$  away from the long-time steady state for some time-intervals \cite{mukherjee2015efficiency}. The distinct properties of Markovian and non-Markovian dynamics raises questions regarding the existence and characteristics of DTCs in the non-Markovian regime.  For example, in a non-Markovian dynamics with a continuously varying $\kappa(t)$, the dissipative dynamics slows down significantly close to $\kappa(t) \to 0$, which may be beneficial for the stabilization of a DTC phase.
Below we focus on the specific setup of a Dicke model in the presence of a Jaynes-Cummings-like dissipation, to show that indeed, the behavior of DTC in the non-Markovian regime can be significantly different from that seen in the Markovian regime. 

\subsection{Modulated open Dicke model}
\label{secIIB}

We consider an open Dicke model comprising $N$  two-level atoms in a cavity, which is described by the Hamiltonian \cite{dickemodel} 
\ba
 {\hat{H}(\lambda) = \omega \hat{a}^\dagger \hat{a} + \omega_{0}\hat{J}_z  + \frac{2 \lambda_t}{\sqrt{N}} (\hat{a} + \hat{a}^\dagger) \hat{J}_x}.
 \label{eqHamil}
\ea
Here $a$ and $a^\dagger$ are respectively the Bosonic annihilation and creation  operator for the photons, $J_\mu = \sum_{i=1}^{N}  \sigma_i^{\mu}$, where $\sigma_i^{\mu}$ denotes the Pauli matrix corresponding to the $i$th spin along the $\mu = x, y, z$ axis, $\omega$ is the frequency of photon field, $\omega_0$ represents the transition frequency of the two level atoms, and $\lambda_t$ is the atom-photon coupling strength, which we consider to be time-dependent in general. 

The setup considered here possesses a $\mathcal{Z}_2$ symmetry; the Hamiltonian \eqref{eqHamil} commutes with the parity operator $ P = e^{i\pi (\hat{a}^\dagger \hat{a}+\hat{J_z}+N/2)}$, such that $\hat{H}(\lambda)$ remains invariant under the transformation $\hat{a}\rightarrow -\hat{a}, $ and $ \hat{J_x}\rightarrow -\hat{J_x} $. However, in the thermodynamic limit $N \to \infty$,  the $\mathcal{Z}_2$ symmetry is spontaneously broken leading to  superradiant phase transition at the critical value of  $\lambda = \lambda_c =  \frac{1}{2} \sqrt{(\omega_0/\omega) (\omega^2 +\kappa_0^2/4)}$  \cite{DickeHEPP1973360,gong18dtc}.

In order to study the possibility of time-translational symmetry breaking in the presence of non-Markovian dynamics, we now consider a phenomenological model, wherein the above atom-photon setup (see Eq. \eqref{eqHamil}) evolves in the presence of a dissipative environment, following the master equation \eqref{eqLgen}.
In analogy with the Jaynes-Cummings model describing the dynamics of a two-level system coupled to a bath comprising Bosonic field modes characterized by a Lorentzian spectral function \cite{Breuer}, we consider here a 
$\kappa(t)$ given by 
\begin{equation}
      { \kappa (t) } =  \begin{cases}
      \frac{2 m \kappa_0\sinh(td/2)}{d \cosh(td/2)+m \sinh (td/2)} & \text{$|\kappa (t)|< \kappa_{\rm{max}}$}\\
      \kappa_{\rm{max} }& \text{ $|\kappa(t)|\geq \kappa_{\rm{max}}$}.\\
    \end{cases}  
    \label{eqkappa}
    \end{equation}
Here $d = \sqrt{m^2-2m \kappa_0}$, $\mathcal{L}[\hat{a}]\hat\rho =\hat{a} \hat{\rho} \hat{a}^{\dagger} - \frac{1}{2} \{{\hat{a}^{\dagger} \hat{a},\hat\rho}\}$ (see Eq. \eqref{eqLgen}), and $\kappa_{\rm{max}} > \kappa_0$ is a parameter which can be tuned to control the maximum possible rate of dissipation.
In case of the  Jaynes-Cummings model, $\kappa_{\rm max} \to \infty$, $m$ denotes the spectral width of the bath, while $\kappa_0$ is related to the system-bath coupling strength \cite{Breuer}. 

The above form of $\kappa(t)$ (Eq. \eqref{eqkappa}) allows us to tune between the Markovian  regime  ($\kappa_0 < m/2$) in which case $\kappa(t) > 0~\forall~t$, and the non-Markovian regime ($\kappa_0 > m/2$), in which case $\kappa(t)$ assumes an oscillatory form realized by replacing $\sinh$ ($\cosh$) by $\sin$ ($\cos$) in Eq. \eqref{eqkappa}, and can can take negative values for some time intervals {\color{blue}(see Sec. \ref{secIInM})} \cite{Breuer, mukherjee2015efficiency}.

Following Ref. \cite{gong18dtc}, one can use Eq. \eqref{eqLgen} to arrive at time-evolution equations for the scaled variables  $x=\langle \hat{a}+\hat{a}^{\dagger}\rangle/\sqrt{2N\omega}$, $p= i \langle \hat{a}-\hat{a}^{\dagger}\rangle/\sqrt{2N\omega}$, and $\bold{j}=(j_x,j_y,j_z)$ with $j_\mu =\langle \hat{J_\mu}\rangle/N$,
 in thermodynamic limit of $N \gg 1$ (see Appendix \ref{appEe}).
 


\subsubsection{Markovian regime}
\label{secIIM}

We start by focusing on the extreme Markovian limit of $m \gg \kappa_0,~ m \to \infty$, in which case $\kappa(t)$ reduces to the time-independent form $\kappa(t) \approx \kappa_0$ for all times. 
For a time-independent $\lambda_t = \lambda_0$, the above setup has two-symmetry broken steady states $\rho_{ss}$ and $\rho_{ss}^{\prime}$ (see Appendix \ref{appEe}) \cite{dimer07proposed, gong18dtc}

In order to study the emergence of time-translational symmetry breaking in the above system, we introduce a period doubling dynamics aimed at periodically evolving between $\rho_{ss}$ and $\rho_{ss}^{\prime}$. To that end,  we apply a periodically modulated $\lambda_t$ of the form
 \begin{equation}
      { \lambda_{t+T} = \lambda_t } =  \begin{cases}
      \lambda_0 & \text{$0 \leq t < T/2$}\\
      0 & \text{ $T/2 \leq t < T$}\\
    \end{cases}  
    \label{eqlambda}
    \end{equation}
   where, 
 $T = \frac{2\pi}{\omega_T}$ and $\lambda_0 > \lambda_c$ \cite{gong18dtc}. In the resonant case $\omega = \omega_0$ and in absence of dissipation ($\kappa_0 = 0$), the above form of $\lambda_t$ results in the parity operation $\hat{P} = \exp[-i (T/2)H(0) + i \pi N/2]$ during the second half period $T/2 \leq t < T$, up to a global phase which we can ignore.

We start with the system in one of the symmetry broken steady states $\rho_{s}$ at the start of a time period. The system stays at $\rho_{s}$ during the first half of the time period, during which time $\lambda_t = \lambda_0$. Thereafter, the modulation in $\lambda$ results in a parity operation $\hat{P}$ during the second half period, which, in the absence of a dissipative bath (i.e., $\kappa_0 = 0$), would take the system to the other symmetry broken steady state $\rho_{s}^{\prime} = \hat{P} \rho_{s}  \hat{P}^{\dagger}$ at the end of the second half period.
Consequently, the above modulation Eq. \eqref{eqlambda} results in $\rho(t) = \rho_s$ ($\rho(t) = \rho_{s^{\prime}}$) at the end of even (odd) number of time periods for $\kappa_0 = 0$, thus giving rise to a DTC behavior. 

In this Markovian regime, the parameter $\kappa(t)$ changes with time for small $t$, finally reaching the value $\kappa(t) = 2 m \kappa_0 /(d + m)$ for long times $t \gg d^{-1}$. However, our numerical analysis shows that the behavior of the system at long times ($t \gg T, d^{-1}$) is independent of $d$.

One can study the robustness of the DTC phase by introducing an error parameter $\epsilon$, defined by $\omega = \left(1 - \epsilon \right)\omega_T; ~ \omega_0 = \left(1 + \epsilon \right)\omega_T$.  As shown in Ref. \cite{gong18dtc}, the DTC order is robust to small values of $\epsilon$, while a larger $\epsilon$ may result in destruction of the same.

\begin{figure}
\begin{center}
     \includegraphics[width= 9cm,height=5.3cm]{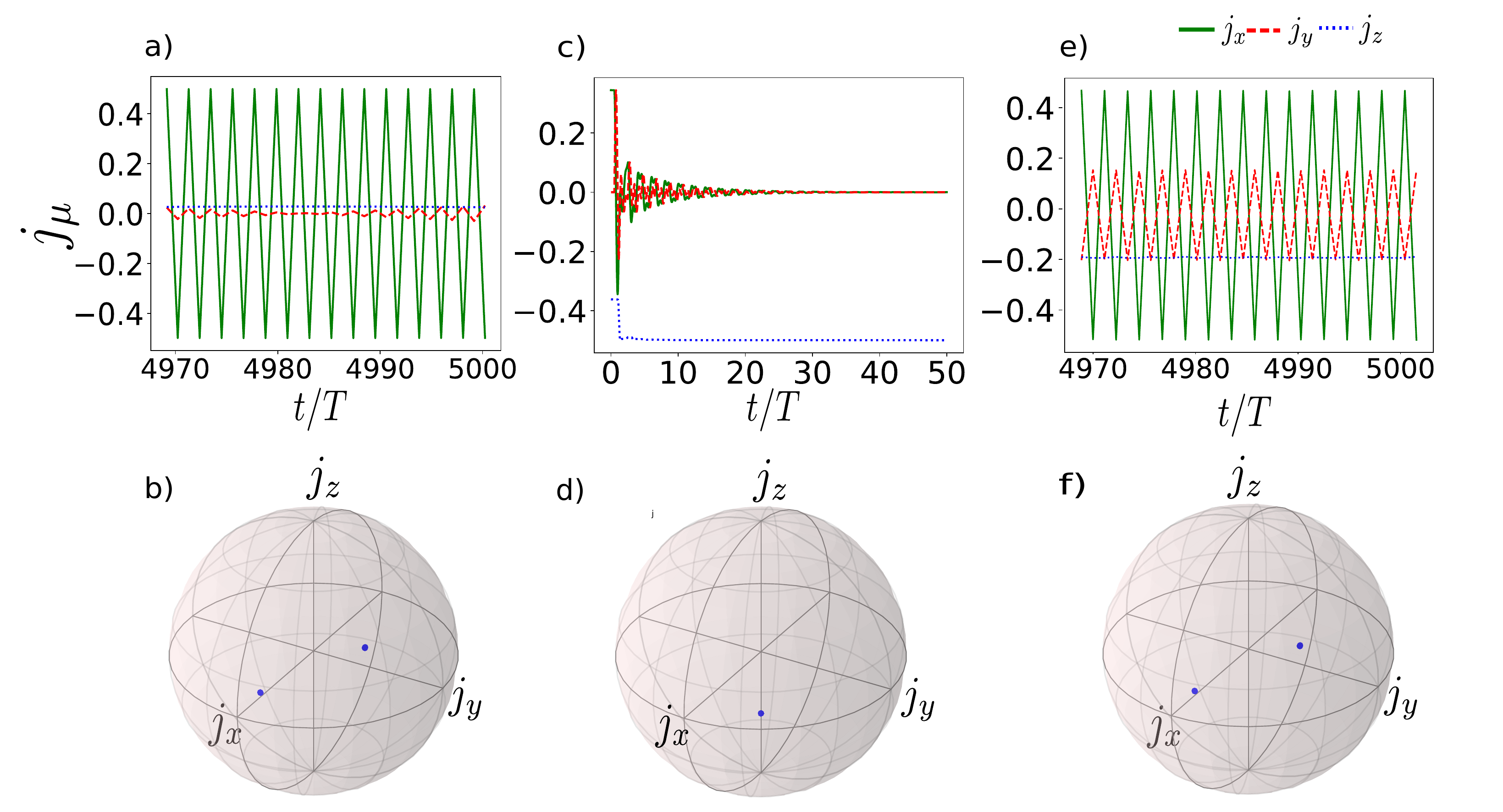}
    \caption{The mean magnetizations $j_x$, $j_y$, $j_z$ and the corresponding Bloch sphere representations are shown for the Markovian with $m \to \infty$ (a, b, c, d) and non-Markovian (e, f) regimes. We get a DTC phase for (a) small $\kappa_0 = 0.05$, represented by two dots in the (b) Bloch sphere. In contrast, a TISS is obtained for large $\kappa_0$ (c, d) in the Markovian regime. The DTC phase is preserved for intermediate values of $\kappa_0$ in the non-Markovian regime (e, f). Here $\omega_T=1$, $T=2\pi$, $\epsilon=0.02$, $\lambda_0=1$, a) $\kappa_0 = 0.05$, c) $\kappa_0=2.7$ c)  $\omega_T=d/2$, $\kappa_0 = 2.7 $ and $m = \kappa_0/4$, $T=2\pi/\omega_T$. }
    \label{fig:comparison}
\end{center}
\end{figure}

Let us now focus on the behavior of the DTC order with increasing values of the  dissipator coefficient $\kappa_0$. The parameter $\kappa_0$ sets the time-scale of thermalization $\tau_{th} \sim \kappa_0^{-1}$. 
Consequently, a low $\kappa_0$ (i.e., $\tau_{th} \gg T$) may be expected to facilitate the emergence of DTC (see Fig.  \ref{fig:comparison}a), as can also be verified in the Bloch sphere representation of the stroboscopic dynamics, where a DTC corresponds to two distinct states, shown by the blue dots in Fig. \ref{fig:comparison}b. On the other hand, in the limit of large $\kappa_0$ (i.e., $\tau_{th} \ll T$), the dissipative mechanism may be expected to dominate the dynamics, thus leading to destruction of the DTC phase, and emergence of a TISS (see Figs. \ref{fig:comparison}c - \ref{fig:comparison}d) \cite{zhu19dicke}.   Therefore it is crucial to study scenarios where a DTC order might be robust over a wide range of parameter values. Below, we go beyond the Markovian approximation to show that remarkably, non-Markovian dynamics may allow us to achieve the above aim, thus resulting in DTC which is robust to both $\epsilon$ and $\kappa_0$.

\subsubsection{Non-Markovian regime}
\label{secIInM}
 
In this section, we focus on the non-Markovian regime, obtained for $\kappa_0 > m/2$. In this case $\kappa(t)$ assumes the form (see Eq. \eqref{eqkappa})
\begin{equation}
      { \kappa (t) } =  \begin{cases}
      \frac{2 m \kappa_0\sin(t|d|/2)}{|d| \cos(t|d|/2)+m \sin (t|d|/2)} & \text{$|\kappa (t)|< \kappa_{\rm{max}}$}\\
      \kappa_{\rm{max}} & \text{ $|\kappa(t)|\geq \kappa_{\rm{max}}$}.\\
    \end{cases}  
    \label{eqnonmarkov}
    \end{equation}
This regime corresponds to a periodically varying $\kappa(t)$ with a time period $T_{\rm NM} = 4\pi/|d|$; $\kappa(t) < 0$ for some time intervals, which eventually results in the so called information back-flow in the system \cite{Breuer, bylicka17constructive}. Here for simplicity we consider a periodically modulated $\lambda_t$ with time period $T = T_{NM}$ (see Eq. \eqref{eqlambda}).

\begin{figure*}[]
    \centering
 
    \includegraphics[width= \linewidth]{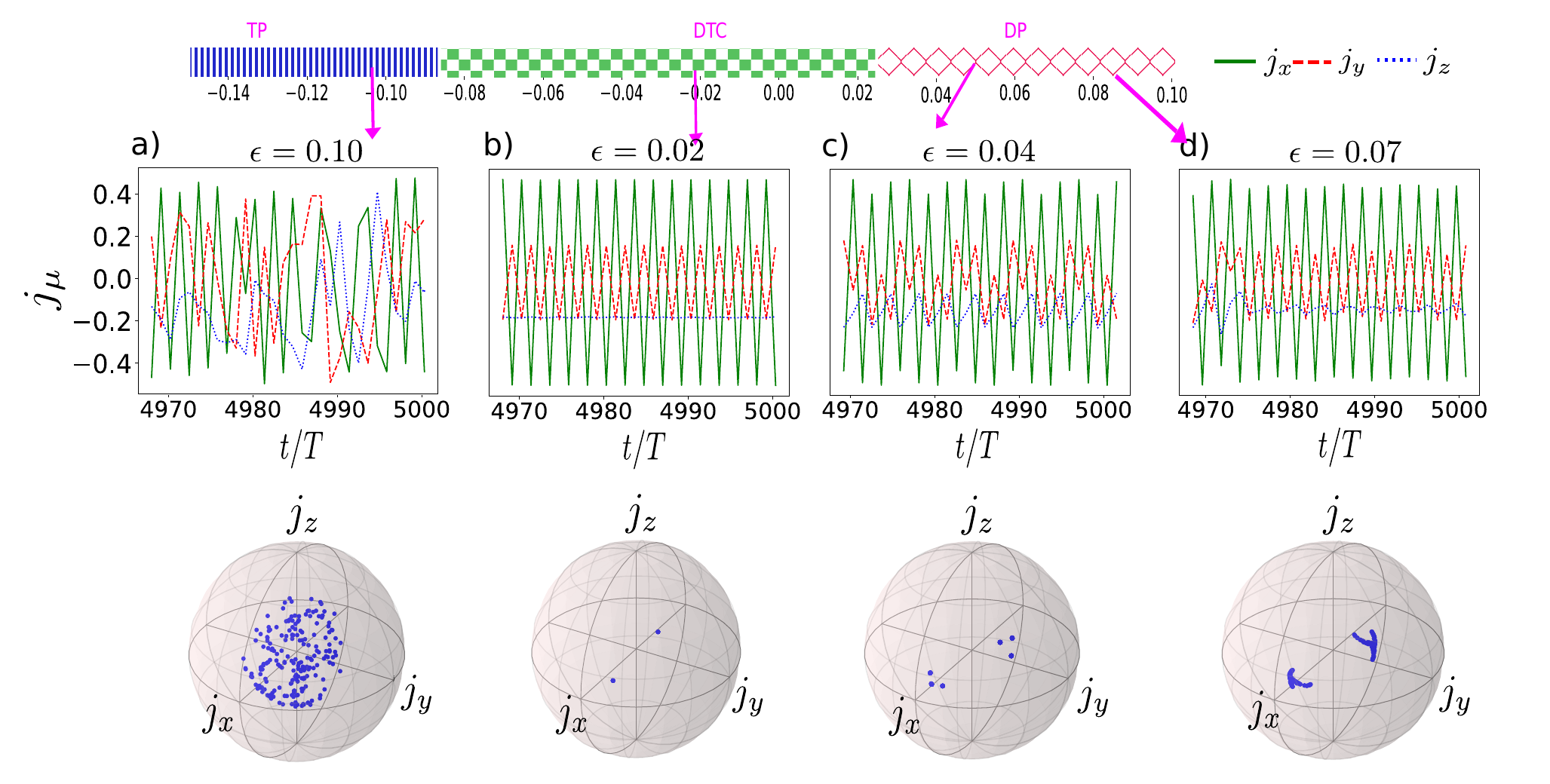}
    \caption{Stroboscopic dynamics (a,b,c,d),  of the atomic average angular momentum $j_{\mu}$ for atom-photon coupling $\lambda = 1$ and time dependent photon-loss rate $\kappa (t)$.   As $\epsilon$  is varied, various dynamical phases emerges over long periods of time.  Only the last 30 periods is shown in the figure. Last 200 periods (blue) are projected onto the Bloch sphere. Here $ \kappa_0 = 2.7$ and $\kappa_{\rm{max}}=5 $.}
    \label{fig:dynamical map}    
\end{figure*}

\begin{figure*}[]
    \centering
    
    \includegraphics[width= \textwidth]{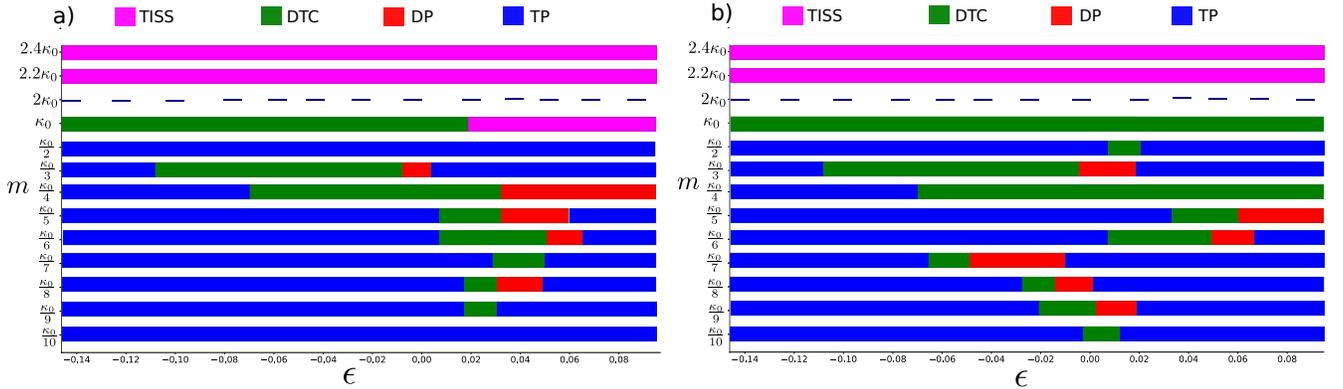}
    \caption{Phase diagram representing TISS, DTC, other dynamical
phases (DP) and thermal phase (TP) as functions of $m$ and  and $\kappa_0=2.7~ a) \kappa_{\rm{max}}=5 , b) \kappa_{\rm{max}}=3$ ,
the dashed line at $m=2\kappa_0$, represents the transition line
from NM to M regime.}
    \label{fig:phase D}    
\end{figure*}

In contrast to the behavior reported for Markovian dynamics \cite{gong18dtc} where DTC phase is present only for small $\kappa_0$ (see Figs. \ref{fig:comparison}a - \ref{fig:comparison}d), numerical analysis shows that  non-Markovian dynamics makes DTC more robust against $\kappa_0$, as signified by the presence of time-crystalline order for intermediate values of $\kappa_0$ and $\kappa_{\rm max}$; this is verified both for the Markovian limit with time-independent $\kappa(t) = \kappa_0$, obtained for $m/\kappa_0 \to \infty$ (see Fig. \ref{fig:comparison}), as well as for finite values of $m/\kappa_0$ (see Figs. \ref{fig:dynamical map} and \ref{fig:phase D}).  This robustness of the DTC phase w.r.t. $\kappa_0$ may be attributed to the periodically varying $\kappa(t)$ in the non-Markovian regime, even at long times \cite{Breuer, hsieh19non}. However, we note that the DTC phase is replaced by a TISS or a thermal phase for large $\kappa_{\rm{max}}$ (see Appendix \ref{apptiss}).

In addition to robustness w.r.t. large $\kappa_0$, the DTC phase shows resilience even in the presence of a non-zero detuning parameter $\epsilon$. Analogous to that seen in the  Markovian regime \cite{gong18dtc},  as we vary $\epsilon$, several  dynamical phases emerge, viz. thermal (Fig. \ref{fig:dynamical map}a),  DTC (Fig. \ref{fig:dynamical map}b), sextet (Fig. \ref{fig:dynamical map}c) and limit cycle (Fig. \ref{fig:dynamical map}d) for these parameters.

In order to have a deeper understanding of the effect of non-Markovian dynamics on the presence of DTC, we study the behavior of the system for different values of $\epsilon$ and $m/\kappa_0$. The results are summarized in the phase diagram Fig. \ref{fig:phase D}. 
We navigate between the Markovian ($m / \kappa_0 > 2$) and the non-Markovian ($m/\kappa_0 < 2$) regimes by varying $m$, for a constant $\kappa_0$. Interestingly, as seen in Fig. \ref{fig:phase D}, one can clearly distinguish the Markovian and non-Markovian regimes from the  stark difference in behavior of the system across the transition.  
 For a large enough $\kappa_0$ ($\kappa_0 \gg T^{-1}$),  the system approaches a TISS in the Markovian regime. However, varied dynamical phases, including DTC, appear on undergoing   transition to the non-Markovian regime at $m=2\kappa_0$ (highlighted by a dashed  line in Fig. \ref{fig:phase D}). Interestingly,  the DTC phase shows the most robustness w.r.t. $\epsilon$ close to the Markovian to non-Markovian transition line $m = 2\kappa_0$. This might be owing to small values of the parameter $d$ for $m \to 2\kappa_0^{-}$, which results in a slowly varying $\kappa(t)$, and an effectively longer time scales $\sim~d^{-1}$. As we move away from the transition line (i.e., $m \ll 2\kappa_0$), we find rich dynamical phases,  including DTC which shows period doubling (see  Fig. \ref{fig:dynamical map} b), sextet which shows periodic behavior with a time period of $6T$ (see  Fig. \ref{fig:dynamical map} c; the corresponding Bloch sphere shows $6$ dots as the system oscillates between $6$ stable solutions  as we vary $\epsilon$), and limit cycle where the system oscillates between two periodic orbits (see Appendix \ref{applc}). Other dynamical phases characterized by different time periods are shown in Fig. \ref{fig:dynamical map} d) and in the red shaded regimes in the Fig. \ref{fig:phase D}. In addition, we get thermal phases, characterized by irregular trajectories and spins randomly distributed on the Bloch sphere  (see  Fig. \ref{fig:dynamical map} a). 
 However, we note that this stark difference between Markovian and non-Markovian regimes vanishes for small $\kappa_0$, in which case varied dynamical phases may exist for different values of $\epsilon$ in the Markovian regime as well.


\subsubsection{Non-Markovian regime with noisy aperiodic $\kappa(t)$}
\label{secRnM}

We next focus on the question whether the  DTC phase in the non-Markovian regime is stable in the presence of random noise in $\kappa(t)$. To this end, we consider an aperiodic $\kappa(t)$, realized by introducing random fluctuations in the dissipator coefficient, given by: 
\ba
\kappa(t)' =\kappa(t) + a_0 f(t). 
\label{eqap}
\ea
Here $f(t)$ ($-1 \leq f(t) \leq 1$) is a random function applied at every time step, while $a_0$ gives the strength of the fluctuations.
\begin{figure}[H]
    \includegraphics[width=\linewidth]{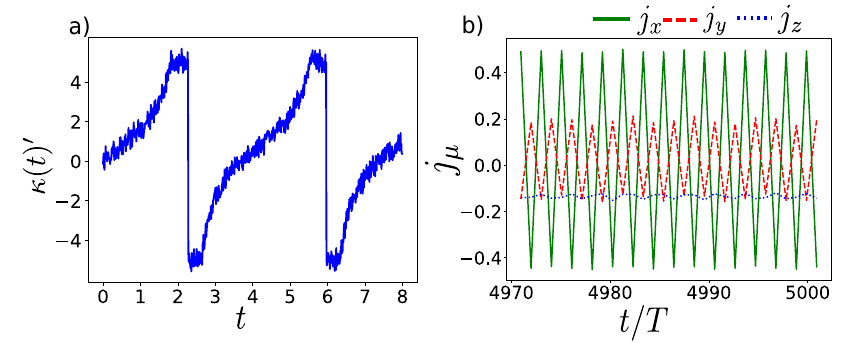}
    \caption{Figure showing (a) an aperiodic $\kappa(t)$, formed through the introduction of random fluctuations in the dissipation (see Eq. \eqref{eqap}), and (b) the corresponding DTC phase. Here $\kappa_0 = 2.7, m=\kappa_0/4,\epsilon=0.03$,$a_0=0.5$ and $\kappa_{\rm{max}}=5$.}
    \label{fig:random}
\end{figure}
As shown in Fig. \ref{fig:random}, the DTC order persists in this case, thereby showing the robustness of DTC phase in the non-Markovian regime, in the presence of intermediate values of $\kappa_0$, non-zero detuning error $\epsilon$ as well as random noise $f(t)$ in the dissipator coefficient. 

\section{Conclusion}
\label{secIII}

We study the emergence of DTC in the presence of non-Markovian dynamics by considering an open Dicke model with Jaynes-Cummings-like dissipation $\kappa(t)$ (Cf. Eq. \eqref{eqkappa}). Our analysis shows that in contrast to Markovian dynamics, the DTC phase is  more robust to a wide range of parameter values in the non-Markovian regime ($m < 2\kappa_0$), which may be attributed to periodically varying $\kappa(t)$ even at long times. We  present a dynamical phase diagram w.r.t. $\epsilon$ and $m$ (Cf. Fig. \ref{fig:phase D}). The transition from Markovian to non-Markovian regime at $m = 2\kappa_0$ is marked by TISS for all values of $\epsilon$ in the Markovian regime for large enough $\kappa_0$, which changes to varied dynamical phases, including DTC, in the non-Markovian regime. Furthermore, the DTC phase emerging for $m$ close to $2\kappa_0$ shows substantial robustness w.r.t. $\epsilon$, while this robustness is reduced, even though does not vanish, as we move deeper into the non-Markovian regime of $m < 2\kappa_0$. Furthermore, we verify the presence of robust DTC order in the non-Markovian regime even for aperiodic $\kappa(t)$, realized by introducing a random noise in the dissipator coefficient. 
Our analysis involves different time scales, which we tabulate below in Table \ref{table}.

We note that the results presented here shows the possibility of using DTC as a probe for non-Markovian dynamics at long times, for intermediate strengths of the dissipator coefficient $\kappa_0$. In the absence of an external modulation, the steady-state of a dynamics may be identical for the Markovian and the non-Markovian dynamics \cite{Breuer, mukherjee2015efficiency}. Consequently, in general a probe for differentiating between the Markovian and non-Markovian regimes is applicable only for short times, before the system reaches the steady-state, which can be highly challenging in cases of short thermalization times. In contrast, as discussed here, the long-time state reached by a system can change drastically in the presence of a periodic modulation; for intermediate values of $\kappa_0$ we get a TISS in case of Markovian dynamics, which changes to a DTC phase for some ranges of $\epsilon$ in the non-Markovian regime (see Figs. \ref{fig:phase D}). This may provide us with a novel way of estimating the nature of the dynamics even at long times, when conventional probes acting in absence of periodic modulation will fail. 

We expect the dissipative DTCs studied here can be realized experimentally in currently existing setups. For example, the presence of DTCs has been verified experimentally in different platforms, both in closed quantum systems, such as in ion traps \cite{zhang17a}, and in the presence of  dissipation, for example in optical cavities \cite{taheri22all,kebler2021observation}. Experimental studies of DTCs in the presence of different forms of dissipation would require control over bath spectral functions, which for example can be achieved through the introduction of filters \cite{naseem20minimal}.

\begin{widetext}
\begin{center}
\begin{table}[H]
\caption{{Time scales}} \label{table} 
\begin{tabular}{|l|l|l|l|}
\hline
Time scales                                                               & \begin{tabular}[c]{@{}l@{}} Markovian (M) \\ regime \end{tabular} & \begin{tabular}[c]{@{}l@{}} Non Markovian(NM) \\regime \end{tabular} & Remarks                                                                                                \\ \hline
\begin{tabular}[c]{@{}l@{}}Modulation period of $\lambda_t$\end{tabular}              & $T$     & $T$       & \begin{tabular}[c]{@{}l@{}}In the DTC phase the system shows oscillations with \\ a time period $2 T$ \end{tabular} \\ \hline
\begin{tabular}[c]{@{}l@{}}Thermalization time $\tau_{\rm th}$\end{tabular} & $\tau_{\rm th}\sim \kappa_0^{-1} $     & $\tau_{\rm th}\sim \kappa_0^{-1}$        & \begin{tabular}[c]{@{}l@{}}DTC phase is preserved for intermediate values of $\kappa_0$\\ ($\kappa_0 \gtrsim T^{-1}$) in the NM regime.\end{tabular} \\ \hline
\begin{tabular}[c]{@{}l@{}}Time period of information \\ back flow $\equiv$ time period of $\kappa(t)$ \end{tabular}          & NA                 & $T_{NM} = 4\pi/|d|$        & \begin{tabular}[c]{@{}l@{}} In the NM regime we have taken $T_{\rm NM} = T$  for simplicity \end{tabular}                                \\ \hline
\end{tabular}
\end{table}
\end{center}
\end{widetext}

Our results show that non-Markovian dynamics can be highly relevant for time-translational symmetry breaking in many-body open quantum systems, and can show features distinct from those seen in the Markovian regime. This also raises open questions regarding the fate of DTCs in more generic forms of non-Markovian dynamics, and regarding the possible role played by information back-flow in the formation of DTCs.

\section*{ACKNOWLEDGEMENTS}

B.D. acknowledges support from Prime Minister Research Fellowship (PMRF).  V.M. acknowledges support from Science and Engineering Research Board (SERB) through MATRICS (Project No.
MTR/2021/000055) and Seed Grant from IISER Berhampur. V.M. also acknowledges Rosario Fazio and Manabendra Bera for helpful discussions. 
 
\appendix

\label{app}
\vspace{1cm}

\section{Evolution equations and steady states}
\label{appEe}

As discussed in Ref. \cite{gong18dtc}, one can use Eq. \eqref{eqLgen} to arrive at the following time-evolution equations for the scaled variables  $x=\langle \hat{a}+\hat{a}^{\dagger}\rangle/\sqrt{2N\omega}$, $p= i \langle \hat{a}-\hat{a}^{\dagger}\rangle/\sqrt{2N\omega}$, and $\bold{j}=(j_x,j_y,j_z)$ with $j_\mu =\langle \hat{J_\mu}\rangle/N$,
 \ba
 \frac{d\textbf{j}}{dt} &=& (-\omega_0 \textbf{e}_z +2\lambda_t \sqrt{2\omega} x \textbf{e}_x )\times \textbf{j}\\ \nonumber
 \frac{dx}{dt} &=& p-\frac{\kappa(t)}{2} x\\ \nonumber
 \frac{dp}{dt} &=& -\omega^2x -\frac{\kappa(t)}{2} p -  2\lambda_t \sqrt{2\omega}x j_x.
 \label{diffeq}
 \ea
 in the thermodynamic limit for $N \gg 1$.

 For a time-independent $\lambda_t = \lambda_0$, the setup considered here has two-symmetry broken steady states $\rho_{ss}$ and $\rho_{ss}^{\prime}$ \cite{dimer07proposed,gong18dtc}
\ba
j_{x}&=&\pm \sqrt{1-\frac{\lambda_c^4}{\lambda^4}},~~
j_y=0, ~~
j_z = \frac{-\lambda_c^2}{\lambda^2}\non\\ 
x &=& \mp \lambda \frac{\sqrt{2\omega(1-\frac{\lambda_c^4}{\lambda^4})}}{\omega^2+\kappa_0^2/4}, \non\\
p&=& \mp\kappa_0/2 \frac{\sqrt{2\omega(1-\frac{\lambda_c^4}{\lambda^4})}}{\omega^2+\kappa_0^2/4}.
\label{solution}
\ea

\section{Dynamics for large $\kappa_{\rm{max}}$}
\label{apptiss}

\begin{figure}[H]
    \includegraphics[width=\linewidth]{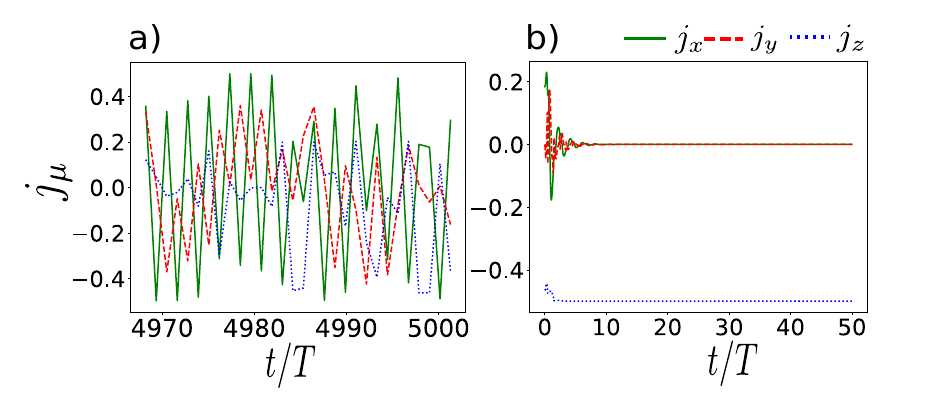}
    \caption{Stroboscopic dynamics for $\kappa_{\rm{max}}=10$, $\epsilon = 0.02$ , a) $m = \kappa_0$ and b) $m = \frac{\kappa_0}{4}$.}
    \label{fig:phaseap}
\end{figure}

In case of large $\kappa_{\rm{max}}$ dissipation dominates the dynamics, such that the system reaches a TISS, as shown in Fig.~\ref{fig:phaseap} a) or a thermal phase, as shown in Fig. \ref{fig:phaseap} b).

\section{Limit cycle}
\label{applc}

\begin{figure}[H]
    \includegraphics[width=\linewidth]{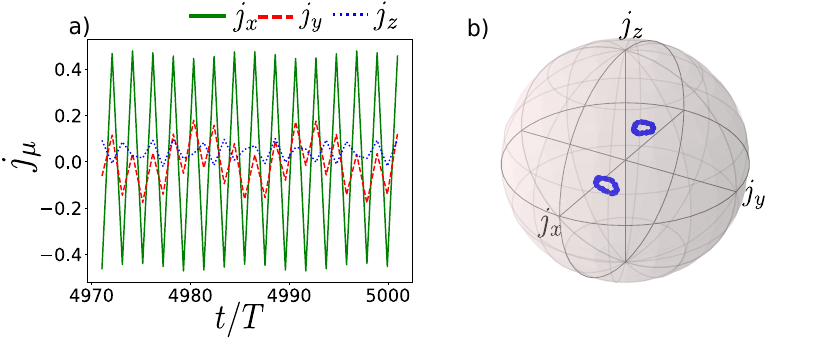}
    \caption{ A limit cycle, Stroboscopic dynamics for $\kappa_{\rm{max}}=3$, $\epsilon = 0.07$ ,  a) $m = \frac{\kappa_0}{5}$.}
    \label{fig:limitcycle}
\end{figure}

As we vary $\epsilon$ the system may oscillate between two limit cycles, as shown in Fig. \ref{fig:limitcycle} (see Sec. \ref{secIInM}).


\end{document}